# A Public Information Precoding for MIMO Visible Light Communication System Based on Manifold Optimization

Hamed Alizadeh Ghazijahani, Mahmoud Atashbar, Guan Yong Liang, *Senior Member, IEEE*, and Zhaojie Yang

*Abstract*—Visible light communication (VLC) is an attractive subset of optical communication that provides a high data rate in the access layer of the network. The combination of multiple input-multiple output (MIMO) with a VLC system leads to a higher speed of data transmission named as MIMO-VLC system. In multi-user (MU) MIMO-VLC, a LED array transmits signals for users. These signals are categorized as signals of private information for each user and signals of public information for all users. The main idea of this paper is to design an omnidirectional precoding to transmit the signals of public information in the MU-MIMO-VLC network. To this end, we propose to maximize the achievable rate which leads to maximizing the received mean power at the possible location of the users. Besides maximizing the achievable rate, we consider equal mean transmission power constraint in all LEDs to achieve higher power efficiency of the power amplifiers used in the LED array. Based on this we formulate an optimization problem in which the constraint is in the form of a manifold and utilize a gradient method projected on the manifold to solve the problem. Simulation results indicate that the proposed omnidirectional precoding can achieve superior received mean power and bit error rate with respect to the classical form without precoding utilization.

*Index Terms*—VLC, MIMO, Precoding, Power efficiency, Public information, Optimization

## I. Introduction

VISIBLE light communication is one of the attractive optical communication systems that utilize the visible region of optical spectrum [1]. Due to the combination of communication with lighting, VLC is one of the emerging technologies in 6G and regarded as a promising technique to provide internet in indoor wireless access [2]. Generally speaking, VLC has many significant advantages, such as license-free spectrum, high security, high data rates, low cost, and freedom from hazardous electromagnetic radiation [3].

Furthermore, multiple-input multiple-output (MIMO) techniques are used in communications systems to ensure a high data rate and reliability where transmitters and receivers use multiple antennas. [4]. A special case of MIMO systems is multi-user MIMO (MU-MIMO) where the transmitter/receiver is equipped with multiple antennas and the receiver/transmitter consists of multiple users with single or multiple antennas. In this case, a base station (BS) with an antenna array supports multiple mobile stations (MS) simultaneously [5].

Recently, to jointly benefit from the advantages of VLC and MU-MIMO, MU-MIMO-VLC systems have been considered, in which a LED array is used to transmit the downlink signals to multiple users simultaneously [6, 7]. To prevent inter-user interference, the users' associated signals are precoded before transmission [8-17]. This type of precoding is named directional precoding where the precode matrix is designed based on the channel matrix. The authors in [8] have formulated the precoding and power allocation problems and do energy efficiency optimizations for multi-cell rate-splitting multiple access VLC systems.

In [9], C. Wang *et al.* proposed a precoding based on successive interference cancellation (SIC) to optimize the electrical/optical power of each LED and achieve the maximum sum rate. Another research designed a precoding matrix by maximizing mutual information subject to both peak and average power constraints [11]. The performance of MU-MIMO-VLC block diagonalization precoding is discussed in [12]. The results show that inter-user interference is eliminated and the complexity of users' terminals is reduced. Another research on precoding for MU-VLC is reported in [13] in which the confidentiality of users' messages has been considered.

In a MU communication system, each user has its private information. The above-mentioned and similar literature employ precoding techniques to send private information to mitigate the other user's signal by maximizing the received signal strength at the intended user. Applying precoding needs to know the channel state information (CSI) of each user. This is while there is some *public* or *common* information that the transmitter unit broadcast simultaneously for all users in the network. This information includes data for synchronization, medium access control frames, link recovery request, or when

Hamed Alizadeh Ghazijahani is with department of electrical engineering, Azarbaijan Shahid Madani University, Tabriz, Iran (hag@azaruniv.ac.ir)

Mahmoud Atashbar (corresponding author) is with department of electrical engineering, Azarbaijan Shahid Madani University, Tabriz, Iran (atashbar@azaruniv.ac.ir)

Guan Yong Liang is with the school of electrical & electronic engineering, Nanyang Technological University (NTU), Singapore (EYLGuan@ntu.edu.sg)

Zhaojie Yang is with the school of electrical & electronic engineering, Nanyang Technological University (NTU), Singapore (zhaojie.yang@ntu.edu.sg)

an IP address is dynamically assigned to a device [18]. To transmit such information, it is assumed that the user location is not known, so the user location feedback is not needed.

Although some type of common information in the VLC network can be transmitted with a private one simultaneously using rate-splitting multiple access (RSMA) technique [19, 20], in RSMA the concept of common information refers to the common message to be decoded by all users but is intended for a single user, which is highly different from the common message that all users need the information contained in the common message [21] in which we refers public message. In addition, similar to directional precoding, in RSMA CSI is needed in the precoding of both common and private information to prevent inter-user interference.

On the other hand, the public message needs to be broadcast to all possible user locations, so inter-user interference is not a challenge here. In this case, we need an efficient precoding for the transmission signal to balance the received signal over the whole coverage area. This type of precoding is named omnidirectional precoding [22, 23]. The main differences between omnidirectional and directional precoding can be summarized as: 1) The transmit information in directional precoding is private information of that user while in omnidirectional precoding, the transmit information is the public information that all users wish to receive, 2) In directional precoding, the channel vector value between the user and transmitter array is known, this is while in omnidirectional only the model of channel is needed, 3) In the directional precoding, the transmitted energy of LEDs is concentrated in the target point where the user is located there. This is while the idea in omnidirectional precoding (assuming an unknown user location) is to maximize the received energy in all potential user locations. To the best of our knowledge, there is not any study that addresses the design of omnidirectional precoding proportional to MU-MIMO-VLC systems.

In this paper, we present an omnidirectional precoding algorithm for a transmitter LED array in a MU indoor MIMO-VLC system for the transmission of public information in the network. To this end, while the user location is not known, we propose to maximize the achievable rate in user potential locations which leads to maximizing the received mean power at the whole area. On the other hand, to achieve higher power efficiency of the power amplifiers used in the LED array, it is needed that the mean transmission powers of the signals associated with all LEDs be the same. Consequently, we consider this constraint besides maximizing the achievable rate in our problem. This leads to a constrained optimization problem in which the constraint is in the form of manifold. Accordingly, to solve this problem, we propose a gradient method projected on the manifold.

The rest of this paper is organized as follows. The system model is described in section II. Section III proposes our optimization problem. The simulation setup, results, and discussion are presented in section IV. Finally, the paper is concluded in section V.

## II. SYSTEM MODEL

Consider a VLC-MIMO system with a uniform rectangular LED array to broadcast public information to the single photo-diode (PD) equipped users in the coverage area shown in Fig. 1. Assuming that the array of LEDs consists of $M_t$ LEDs in a rectangular structure as $M_t = M_x \times M_y$ on the x-y plane and $d_x$ and $d_y$ are the distances of adjacent LEDs along the *x*-axis and *y*-axis, respectively. Note that in our model, the LED panel is placed on the ceiling with height $D$ from the room's floor. Furthermore, assume that the public information signal vector $\boldsymbol{s} = [s_1, s_2, \ldots s_q]^T$ sent to users, where $s_i$, $q$, and $(.)^T$ indicate the *i*-th transmitted symbol, number of symbols, and transpose, respectively. This signal vector is multiplied by the designed precoding matrix $\boldsymbol{P} \in \mathbb{R}^{M_t \times q}$ to generate LEDs' transmission signal vector $\boldsymbol{v} = \boldsymbol{Ps}$, where $\boldsymbol{v} = [v_1, v_2, \ldots v_{M_t}]^T$ with $v_i$ presents the transmit signal of *i*-th LED located in $(x_i, y_i, D)$. The coordinate of *i*-th LED is

$$x_i = \left\lfloor \frac{i-1}{M_y} \right\rfloor d_x, \; y_i = (i \bmod M_y - 1)d_y, \; i = 1, 2, \ldots M_t \quad (1)$$

in which $\lfloor . \rfloor$ denotes the floor operator sign and 'mod' indicates the reminder of deviation.

In VLC systems, each user is equipped with a PD to receive the transmitted optical signal strength from LEDs array. Thus, the received signal at the *j*-th user in the coordinates $(x_j^u, y_j^u, h)$ will be as follows [11]:

$$r_j = \boldsymbol{h}_j^T \boldsymbol{P} \boldsymbol{s} + n \quad (2)$$

where $\boldsymbol{P}$ is the $M_t \times q$ precoding matrix, $n$ is white Gaussian noise and $\boldsymbol{h}_j$ is the $M_t \times 1$ channel vector. $\boldsymbol{h}_j = [h_{j1}, h_{j2}, \ldots h_{jM_t}]^T$ with $h_{ji}$ is the VLC channel gain between the *i*-th LED and the *j*-th user.

VLC channel models are currently investigated under two categories: deterministic and stochastic models. Deterministic models aim to predict channel characteristics at a specific location of the transmitter and receiver, as well as the surrounding environment, with ray tracing, recursive, and empirical algorithms. In stochastic approaches, the impulse responses of VLC channels are defined by the law of light propagation applied to a specific geometry of transmitter, receiver, and scattered [24]. In this research, we also use deterministic channel model as commonly used in other literature.

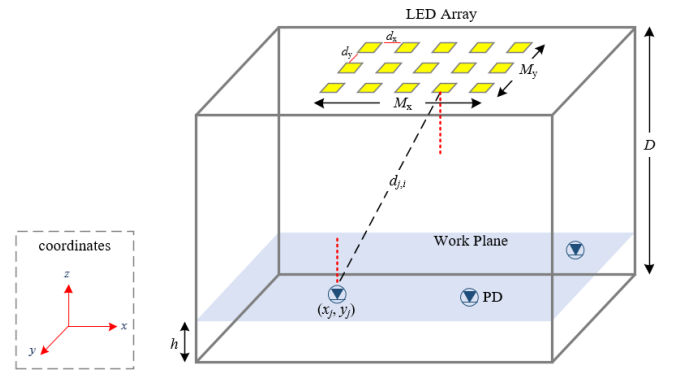

Fig. 1. System model

As the line of sight (LOS) channel of the optical wireless channel contains most parts of the transmitted energy [9], we ignore the non-LOS part in this work. According to the geometry presented in Fig. 1, for the LOS path, $h_{ji}$ is given as [20]

$$h_{ji} = \begin{cases} \frac{A_d(m_l+1)}{2\pi d_{ji}^2}\cos^{m_l}(\phi_{ji})\cos(\theta_{ji})TG & if\ 0 \leq \theta_{ji} \leq \psi_R \\ 0 & if\ \theta_{ji} < \psi_R \end{cases} \quad (3)$$

where $\phi_{ji}$ denotes the emitting angle, $\theta_{ji}$ denotes the incident angle from the $i$-th LED to $j$-th user, $A_d$ denotes the area of the receiver PD, $m_l$ is Lambert's mode number expressing directivity of the source beam, the $T$ denotes the signal transmission coefficient of an optical filter, $\psi_R$ denotes the field of view of the receiver PD, $G$ denotes the concentrator gain, and $d_{ji}$ is the distance between $i$-th LED to $j$-th user PD [25].

### III. PROPOSED OPTIMIZATION

In this paper, the idea is to design an omnidirectional precoding matrix for efficient transmission of public information signals of all users distributed on the coverage area in the VLC network. To cope with this, two limitations are considered. 1) Maximum achievable rate, 2) Keep constant the mean transmission power of array LEDs.

*A. Maximum achievable rate*

According to the channel model described in (2), the mutual information of MIMO-VLC downlink transmission will be as follows [26, 27]

$$I_j = \log\left(1 + \frac{1}{\delta^2}\boldsymbol{h}_j^H \boldsymbol{P}\boldsymbol{P}^H \boldsymbol{h}_j\right) \quad (4)$$

in which $\delta^2$ indicate the variance of noise at the receiver and $(.)^H$ indicates the Hermitian. Besides, it is supposed that the mean power of the elements of the vector $\boldsymbol{s}$ is equal to one. According to (4), maximizing term $\boldsymbol{h}_j^H \boldsymbol{P}\boldsymbol{P}^H \boldsymbol{h}_j$ leads to maximum achievable rate where it is in line with maximizing received mean power at $j$-th user. Since the users can be located at any point of the covered area, so the precoding matrix should be designed in a way that term $\boldsymbol{h}_j^H \boldsymbol{P}\boldsymbol{P}^H \boldsymbol{h}_j$ get maximum for all possible $h_j$ values. For this aim, first, we do sampling from the coverage area, then maximize the average of received mean power (ARMP) at the sampled location points as

$$ARMP = \frac{1}{N_s M_t}\sum_{j=1}^{N_s} \boldsymbol{h}_j^H \boldsymbol{P}\boldsymbol{P}^H \boldsymbol{h}_j \quad (5)$$

with $N_s$ is the number of sample location points. In this equation, the parameter $M_t$ is utilized to normalize the total transmit power at the LED array.

*B. Keep constant the mean transmission power of LEDs in the array*

In a similar way to the radio frequency MIMO transmission system [28], to achieve a higher power efficiency of the amplifiers used to drive the LED array, it is needed that the mean transmission powers of the signals associated with all LEDs be the same. According to the unit mean power assumption of vector $\boldsymbol{s}$ elements, the declared constraint becomes $\sum_{i=1}^{q}|P_{mi}|^2 = 1$, $m = 1,2,\ldots,M_t$ where $P_{mi}$ represents the $(m, i)$-th element of the matrix $\boldsymbol{P}$ with $|P_{mi}| \leq 1$. This constraint can be expressed in the matrix form as follows

$$diag(\boldsymbol{PP}^H) = \boldsymbol{I}_{M_t} \quad (6)$$

in which, $\boldsymbol{I}_{M_t}$ is the $M_t \times M_t$ identity matrix and $diag$ (.) represents a diagonal matrix whose major diagonal elements are equal to the major diagonal elements of the matrix.

Accordingly, the proposed constrained optimization problem to design the precoding matrix $\boldsymbol{P}$ is

$$\begin{aligned}\max_{\boldsymbol{P}} \quad & \sum_{j=1}^{N_s} \boldsymbol{h}_j^H \boldsymbol{P}\boldsymbol{P}^H \boldsymbol{h}_j \\ s.t. \quad & diag(\boldsymbol{PP}^H) = \boldsymbol{I}_{M_t}\end{aligned} \quad (7)$$

As the term $\sum_{j=1}^{N_s} \boldsymbol{h}_j^H \boldsymbol{P}\boldsymbol{P}^H \boldsymbol{h}_j$ is a concave function, by choosing $f(\boldsymbol{P}) \triangleq -\sum_{j=1}^{N_s} \boldsymbol{h}_j^H \boldsymbol{P}\boldsymbol{P}^H \boldsymbol{h}_j$, it transforms to a convex function. Accordingly, the optimization problem forms as

$$\begin{aligned}\min_{\boldsymbol{P}} \quad & f(\boldsymbol{P}) \\ s.t. \quad & diag(\boldsymbol{PP}^H) = \boldsymbol{I}_{M_t}\end{aligned} \quad (8)$$

Recently, geometric solutions are used to solve various optimization problems. One kind of such solutions is manifold-based geometry which is used in constrained optimization problems [29, 30] because of its relative simplicity and optimality. The constraints in constrained optimization problems can be interpreted as isolated points in the space that are in the manifold forms such as Stiefel, Grassmann, Riemannian, etc. Consequently, the optimum points are searched in the space that is inside the manifold.

In this work, we propose a manifold-based method to solve (8). As the constraint in (8) is in the form of Grassmann manifold [29] and $f(\boldsymbol{P})$ is in the form of a quadratic function so is a convex function, we use the gradient method projected on the manifold, in which, matrix $\boldsymbol{P}$ is calculated iteratively as

$$\boldsymbol{P}_{k+1} = \boldsymbol{P}_k + \mu \nabla f(\boldsymbol{P}_k) \quad (9)$$

in which $\boldsymbol{P}_k$ is the $\boldsymbol{P}$ values in $k$-th iteration, $\mu$ is step size, and $\nabla f(\boldsymbol{P})$ is the $M_t \times q$ gradient matrix of $f(\boldsymbol{P})$. According to matrix relations on [31], we have:

$$\nabla f(\boldsymbol{P}) = -2\sum_{j=1}^{N_s} \boldsymbol{h}_j \boldsymbol{h}_j^H \boldsymbol{P} \quad (10)$$

In each iteration of the gradient algorithm, to ensure the constraint is established, the resulting matrix $\boldsymbol{P}_{k+1}$ is projected on the manifold. Since the constraint in (8) is in the form of the Grassmann manifold, the projection on the above manifold is

as [29]

$$\boldsymbol{P}_{k+1} \leftarrow \left(diag(\boldsymbol{P}_{k+1}\boldsymbol{P}_{k+1}^H)\right)^{-\frac{1}{2}} \boldsymbol{P}_{k+1}. \quad (11)$$

The iteration is continued until the difference of $f(\boldsymbol{P}_k)$ goes below a determined small value $\varepsilon$ for two successive iterations to satisfy the coverage condition, as $|f(\boldsymbol{P}_{k+1}) - f(\boldsymbol{P}_k)| < \varepsilon$. The steps of the proposed algorithm to solve the optimization problem (8) are determined by the projected gradient method on the manifolds presented in Algorithm 1.

Algorithm 1. Solving optimization problem (8)
1- Initialization of matrix $\boldsymbol{P}$
2- Calculate $\nabla f(\boldsymbol{P})$ using (10)
3- Update $\boldsymbol{P}$ as (9)
4- Project $\boldsymbol{P}$ on the manifold based on (11)
5- Repeat steps 2 to 4 to achieve the convergence condition

## IV. SIMULATION, RESULTS, AND DISCUSSION

In this section, we present the simulation setup and results to show how the proposed precoding algorithm for MIMO-VLC satisfies two limitations stated in section II. To this end, we consider optimized ARMP as an evaluation criterion in which optimized ARMP is defined as the value of ARMP based on the designed precoding matrix $\boldsymbol{P}$. To the best of our knowledge, there is not any similar study to design an omnidirectional precoding matrix for MIMO-VLC system, we choose the mean of ARMP parameters over all random precoding matrix $\boldsymbol{P}$ as a reference method to compare the results. We name this as 'classical method'.

In the simulation, we consider a scenario in which room dimensions are 5, 6, and 3 meters for width, length, and height, respectively. A LED uniform rectangular array is supposed to be installed in the center of the room ceiling. Also, the users can be placed at all possible locations on the floor of the room and each of them receives the VLC signal emitted from all LEDs.

As mentioned in section 3, to aim for maximum achievable rate, we need to do sampling from all possible locations of users, therefore, in our simulations, the floor of the room is sampled uniformly with a distance of 0.1 in both axes. The other simulation parameters are chosen as $\varepsilon = 10^{-4}, q = 10$, and $\mu = 10^8$.

### A. Convergence of optimization problem

In the first simulation, the convergence of the gradient method projected on the manifold in solving the proposed optimization problem is investigated. In this way, a 3×3 LED rectangular array with a distance of 0.02 m between adjacent elements is considered at the ceiling and the PD of users is located on a flat surface, named work plane, with height 1m from the floor. The $f(\boldsymbol{P})$ value is calculated in each iteration. The result shown in Fig. 2 indicates that the cost function is converged in 5-th iteration. The resultant precoding matrix in 5-th iteration is as (12) showing that the constraint of proposed optimization problem presented in (14) is satisfied.

### B. Number of LED array elements

In the second simulation, the behavior of ARMP versus different LED numbers in the array is investigated. In this simulation the parameters are set as $d_x = 0.01$, $h = 2.5$ m, and $M_t$ the number of LEDs varies from 4 to 64. Figure 3 shows that the ARMP of the proposed method is improved by increasing the number of LEDs, while the classical ARMP is constant over $M_t$ changes, as expected. This is due to the fact that as the elements of $\boldsymbol{P}$ matrix are chosen random and according to the law of large numbers, the ARMP is proportional to the variance of $\boldsymbol{P}$ elements.

Based on the considered parameters in simulation, the ARMP value for the classical method is almost fixed at 5.2e-10 for all $M_t$ values. This is while, for the proposed method the ARMP is 4.6e-9 and 3.3e-8 for $M_t = 9$ and 64, respectively. The received signals from LEDs at each user location sum up linearly in classical method which is not necessarily constructive while, the designed precoding leads to a constructive summation of LEDs' signals in the proposed method. By increase in $M_t$, the

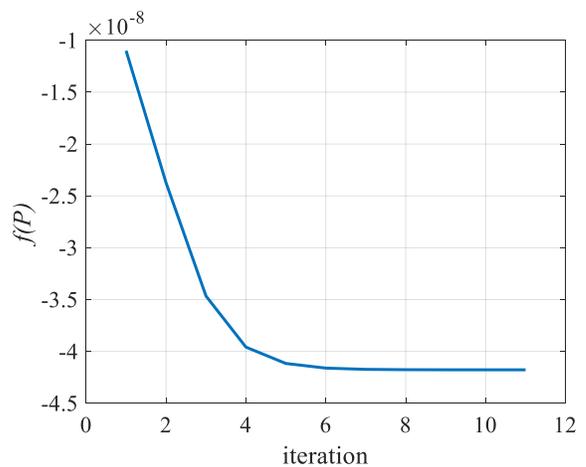

Fig. 2. Convergence of proposed algorithm over iterations

$$\boldsymbol{P} = \begin{bmatrix} 0.485532 & 0.389968 & 0.148727 & -0.28536 & 0.032314 & 0.279723 & -0.33371 & 0.459929 & -0.06347 & 0.320037 \\ 0.063389 & -0.5254 & -0.0869 & 0.357057 & -0.25406 & -0.22072 & 0.185637 & 0.187258 & -0.33472 & -0.53857 \\ -0.29119 & -0.13221 & -0.16466 & -0.47472 & 0.395979 & 0.432586 & 0.102007 & 0.273667 & -0.21922 & -0.40984 \\ 0.031884 & -0.35761 & -0.07127 & -0.35639 & 0.424023 & -0.39309 & 0.287618 & -0.14704 & -0.4282 & 0.342038 \\ 0.243148 & -0.38668 & -0.15495 & -0.31892 & -0.39669 & -0.3259 & -0.21353 & -0.09049 & -0.49624 & 0.319427 \\ -0.37737 & -0.4311 & 0.093535 & -0.22487 & 0.246277 & 0.378553 & 0.145202 & 0.385867 & 0.450226 & -0.1892 \\ -0.45466 & -0.23233 & 0.337457 & 0.352245 & -0.43656 & 0.389548 & 0.377869 & 0.004552 & 0.058183 & 0.113256 \\ -0.46861 & 0.191308 & 0.274291 & -0.32424 & -0.51558 & -0.02095 & -0.12132 & -0.41916 & -0.05102 & 0.322735 \\ 0.087989 & 0.13417 & -0.16224 & -0.12184 & 0.365285 & 0.004531 & 0.209953 & 0.693242 & -0.44237 & 0.281573 \end{bmatrix} \quad (12)$$

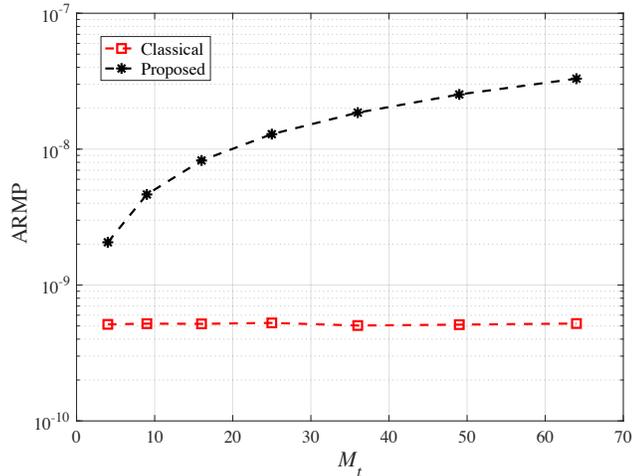

Fig. 3. The ARMP of classical and proposed methods versus number of LEDs in the array with $d_x = 0.05$ m.

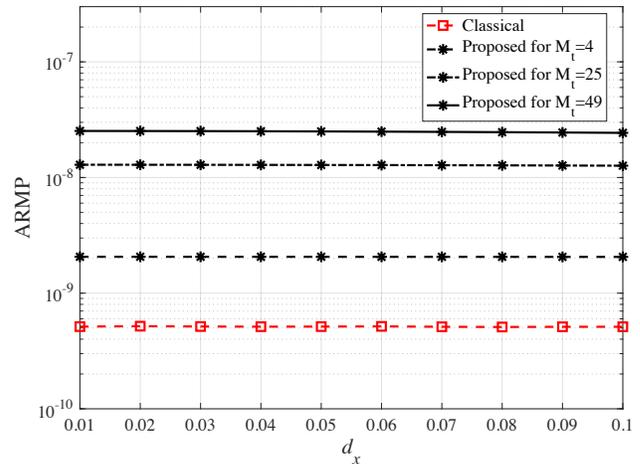

Fig. 4. The ARMP of proposed method under $M_t = \{9, 25, 64\}$ and classical method versus low dynamic range adjacent distance between LEDs in the array.

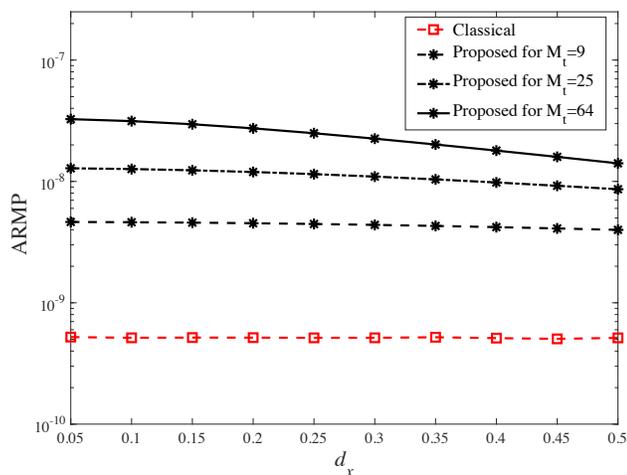

Fig. 5. The ARMP of proposed method under $M_t \in \{9, 25, 64\}$ and classical method versus high dynamic range adjacent distance between LEDs in the array.

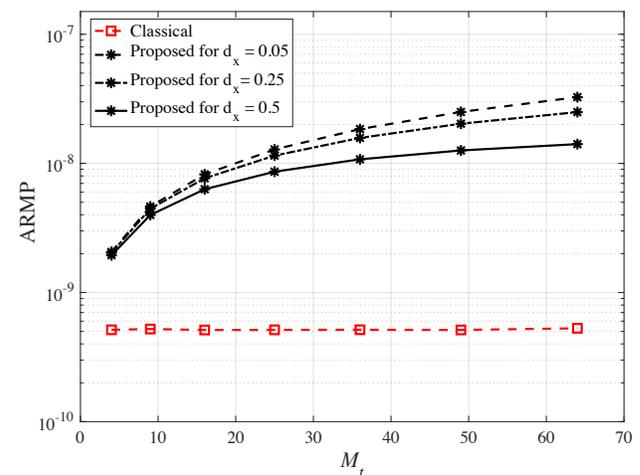

Fig. 6. The ARMP of classical and proposed algorithm versus number of LEDs in the array under some high dynamic range between elements in the array

degree of freedom in the constructive summation is increased and helps to increase optimized ARMP.

### C. Distance between LED array elements

To investigate the impact of distance between LEDs in the array, we repeated the simulation for $d_x = d_y \in \{0.01, 0.02, ..., 0.1\}$. The ARMP versus different distance values is depicted in Fig. 4. In this figure, the simulation results for the proposed method are presented with three different numbers of LEDs in the array. Besides, the ARMP curve for the classical method for any arbitrary $M_t$ is depicted versus $d_x$. As seen, the ARMP values remain unchanged by increasing in $d_x$ in both methods. The constant value of ARMP is due to the low dynamic range of $d_x$. To confirm this matter, the simulation is repeated for a $d_x$ with a high dynamic range where the results are shown in Fig. 5. As seen, for large $M_t$, the ARMP curve of the proposed method falls by an increase in $d_x$ while it is almost constant for small $M_t$'s. This is due to the fact that when the $M_t$ and $d_x$ are concurrently large, the panel of LED array physical length goes expand over the ceiling and this makes the constructive combination of LED signals hard in most points of the work plane. To show the impact of LED numbers, the ARMP versus $M_t$ is depicted in Fig. 6 for some high dynamic ranges between elements in the array. As seen, although by increasing $M_t$, the LED array physical length is running larger, the optimized ARMP has incremental functionality with $M_t$. It has resulted that in large LED array panels, considering both $M_t$ and dx jointly, $M_t$ has a dominant impact on the ARMP of proposed method.

Finally, the performance of proposed and classical methods under different work plane heights is studied. To this end, we vary the work plane from the floor up to the height of 1m. We set dx = 0.05m and $M_t \in \{9, 25, 64\}$ in our simulations. The result is shown in Fig. 7 in which the horizontal axis is the work plane height from the floor. As expected, by moving the work plane from the floor the ARMP for both proposed and classical methods increases. This is due to the fact that by increasing work plane height, the distance between user locations and the LED array decreases which leads to a decrease in channel loss.

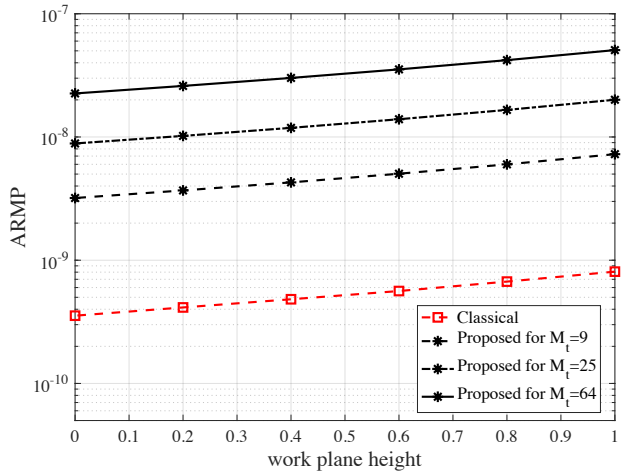

Fig. 7. The ARMP of classical and proposed algorithm versus work plane height under different number of LEDs in the array

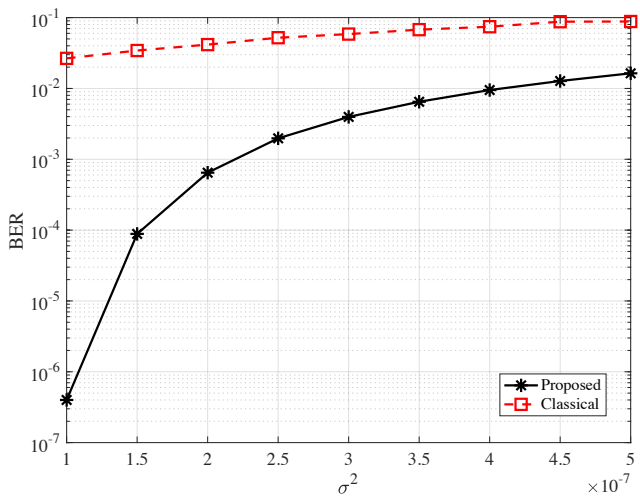

Fig. 8. The BER of proposed and classical methods versus noise variance

### D. Bit error rate analysis

For more investigation of the system performance, the bit error rate (BER) of the proposed method is compared with that of the classical method. Detection of public information at $j$-th user needs the value of the term $\boldsymbol{h}_j^T \boldsymbol{P}$ to be known. It is assumed that a pilot block of $q$ symbols is broadcast at the first so the $j$-th user can estimate the relevant value of $\boldsymbol{h}_j^T \boldsymbol{P}$. Then, the main public information bits are modulated with usual on-off keying scheme then broadcast to all users. At the receiver side, each user detects the public information bits using the Maximum-Likelihood criterium based on the estimation of term $\boldsymbol{h}_j^T \boldsymbol{P}$. The mean BER of 15 users distributed uniformly in the work plane versus noise variance is depicted in fig. 8 for both proposed and classical methods under $M_t = 9$, $d = 0.02$ m. As seen, the BER values of the proposed methods encouragingly outperforms the classical method.

## V. CONCLUSION

In this paper, we proposed an omnidirectional precoding for transmitting the public signals in MU-MIMO-VLC system. For this purpose, we proposed an optimization problem which maximizes the received mean power constrained with equal transmission mean power of LEDs in the array. In our formulation the constraint is in the form of manifold therefore a gradient method projected on the manifold is designed to solve the problem. We considered the ARMP parameter to investigate the performance of the system under varying some simulation parameters such as LED numbers in the array, distance between LEDs, and height of work plane from the floor. Simulation results has shown that the proposed omnidirectional precoding leads to higher ARMP values with respect to the classical method in all simulation scenarios.